\documentclass[apj]{emulateapj}

\shorttitle{Removing Reflections from Imaging Data}
\shortauthors{Slater, Harding, and Mihos}

\begin{document}

\title{Removing Internal Reflections from Deep Imaging Datasets}

\author{Colin T. Slater,\altaffilmark{1} Paul Harding, and J. Christopher Mihos}
\affil{Department of Astronomy, Case Western Reserve University,
  Cleveland, OH.}
\email{colin@astronomy.case.edu, paul.harding@case.edu, mihos@case.edu}
\altaffiltext{1}{now at the Department of Astronomy, University of Michigan}

\begin{abstract}

We present a means of characterizing and removing internal reflections
between the CCD and other optical surfaces in an astronomical camera.
The stellar reflections appear as out-of-focus images and are not
necessarily axisymmetric about the star. Using long exposures of very
bright stars as calibration images we are able to measure the
position, size, and intensity of reflections as a function of their
position on the field.  We also measure the extended stellar
point-spread function out to $1 \degr$.  Together this information can
be used to create an empirical model of the excess light from bright
stars and reduce systematic artifacts in deep surface photometry. 
We then reduce a set of deep observations of the Virgo cluster with
our method to demonstrate its efficacy and to provide a comparison
with other strategies for removing scattered light.

\end{abstract}


\section{Introduction}

Deep wide-field imaging is widely used to study extended, diffuse
objects such as low surface brightness galaxies
\citep{McGaugh95,Sprayberry96,Marshall04}, the diffuse interstellar
medium \citep{Sandage76,Gordon98,Witt08}, the outer disks and faint
stellar features around bright galaxies
\citep{Malin83,Pohlen02,Delgado09}, the diffuse intracluster light in
galaxy clusters \citep{Uson91,Gregg98,Feldmeier04,Gonzalez05,Mihos05},
and the extragalactic background light \citep{Bernstein07}. Accurately
measuring faint surface brightnesses places stringent demands on the
minimization of systematic effects, including large scale flat
fielding, accurate sky subtraction, scattered light from nearby
objects, and internal reflections in the telescope/camera system. All
of these effects, if not treated carefully, can imprint a spatially
varying pattern of light onto the image, which can significantly
contaminate measurements of the low surface brightness astronomical
object being studied.

In particular, internal reflections of bright stars in the field
represent a significant source of this type of contamination. Light
coming to focus can reflect off the CCD and back up the optical path,
reflect again off optical elements such as the dewar window, filter,
or any reimaging optics, and come back down to the CCD. The longer
path length results in multiple out-of-focus stellar images (one for
each reflection) being added to the image. These defocused stars are
essentially extended images of the telescope's entrance pupil, and
have complex spatial structure due to obstructing objects such as the
Newtonian mirror and its support spider. As we show below, the
position of these reflections relative to their central star changes
across the field of view, making characterization and subtraction
difficult. They also complicate the measurement of the extended
stellar PSF, making the subtraction of the low surface brightness
wings of stars problematic. The complex pattern of overlapping
reflections and stellar wings can then plague efforts to accurately
model and subtract sky from the images.

Some of this contamination can be reduced in hardware, through careful
baffling of the telescope to reduce scattered light, minimization of
optical elements to reduce the number of reflecting surfaces, and the
use of aggressive anti-reflective coatings on the optical surfaces to
reduce the intensity of the reflections. A particular example of this
kind of effort is our optimization of Case Western Reserve
University's Burrell Schmidt for deep surface photometry.  This
24/36-inch Schmidt telescope is located at Kitt Peak and was
originally designed for wide field imaging with photographic plates
\citep{Nassau45}. The telescope was subsequently converted to
Newtonian focus with a flat Newtonian mirror and a CCD imaging camera
located on the side of the telescope tube. 

The telescope is particularly well-suited to deep surface photometry
by its original design, and in addition we have made additional
upgrades to increase its sensitivity. The telescope's closed-tube
design and the use of a Newtonian focus severely limits the amount of
stray light that reaches the detector. Recent upgrades to the
telescope include a redesign of the Newtonian mirror and its mounting
structure to reduce vignetting and flexure, an installation of
light-absorbing material to the inside of the telescope tube to reduce
scattered light, the installation of a wide field 4Kx4K CCD
($1.65\degr \times 1.65\degr$ on the sky), a combining of the dewar
window and field flattener to reduce optical surfaces, and the use of
filter-specific anti-reflective coatings on the filter and dewar
window/field flattener (see Appendix A). With these improvements, the
telescope has been able to detect extremely faint structure in the
intracluster light of Virgo \citep{Mihos05}.

However, in practice many of these solutions are generally
unavailable to the observer on a multi-user, multi-instrument
telescope, and even with such solutions in place, faint reflections
still persist. In these situations, software solutions must be
implemented to correct for these reflections in the reduction process,
after the data has been taken.  In this work we present a way of
characterizing and removing these reflections in post-processing. This
provides a cost-effective way to increase the telescopes ability to
image very faint structure. Modeling the reflections in software is
also convenient in that it does not require any modification to the
telescope or other work on the part of the instrument scientist, and
can be implemented by an observer without the aid of the observatory
staff. The only change necessary to the observing program is the
observation of a small number of bright stars in order to characterize
the sizes, intensities, and positions of the reflections. With this
information a generative model of the reflections can be built and
used to remove the scattered light.


The outline of our paper is as follows. In \S\ref{characterizing}, we
characterize the reflections from our Schmidt imaging data and show
that they are as expected for the optical design of the telescope. We
also describe a process for removing them from the images. In
\S\ref{scattering}, we use the reflection subtracted images to
accurately measure the point spread function of the telescope out to
1$\degr$. In \S\ref{sciencereduction}, we examine the scientific
impact of the technique by comparing the results of our deep imaging
of Virgo with and without the full reduction technique in
place. Finally, in \S\ref{summary} we present a step-by-step
``cookbook summary'' of the technique and suggest avenues by which the
technique could be further improved.

\section{Characterizing the reflections}
\label{characterizing}

\begin{figure}
\epsscale{0.5}
\plotone{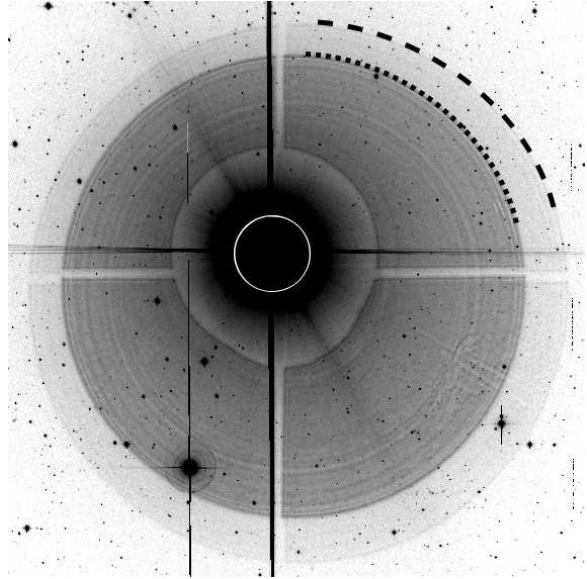}
\epsscale{1}
\caption{An example image of Arcturus, scaled
  logarithmically. The black dashed arc indicates the outer edge of
  the reflection off the top surface of the filter, and the dotted arc
  indicates the edge of the bottom surface filter reflection. The
  solid white circle in the center shows the extent of the dewar
  window reflection, which is not visible under this scaling. The
  image is $41\arcmin \times 41\arcmin$.
\label{arcturus-rings}}
\end{figure}

\begin{figure}
\plotone{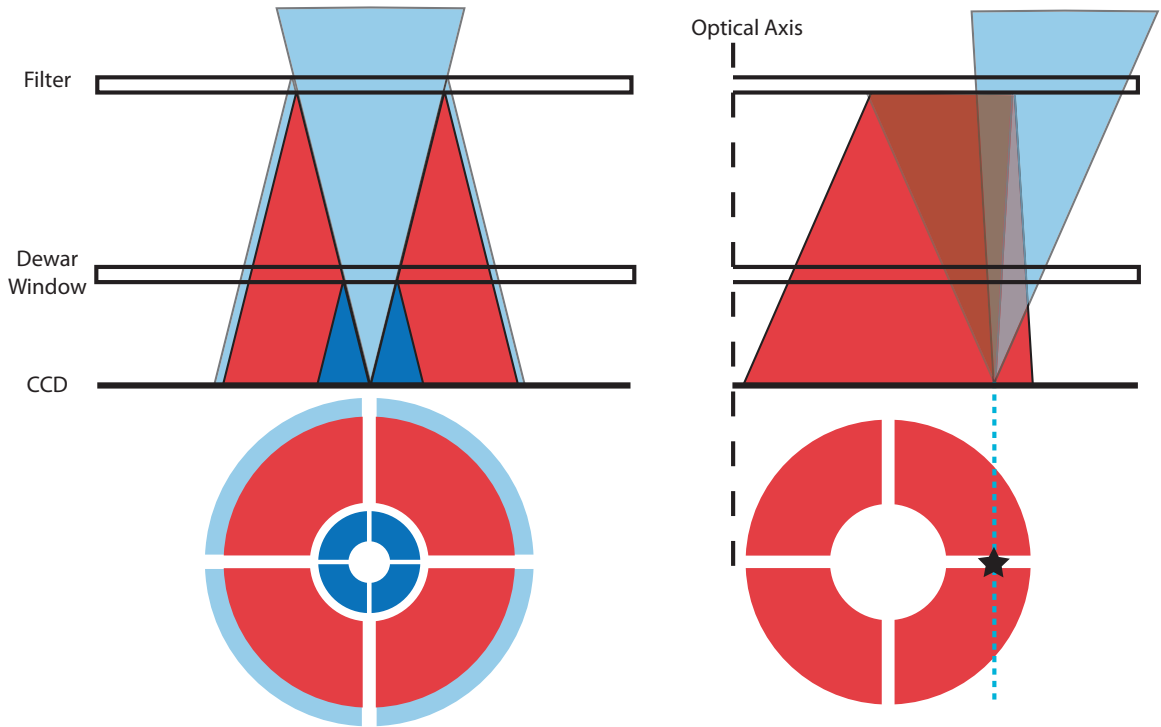}
\caption{Schematic view of the surfaces causing the observed
  reflections. The drawing on the left is for a star in the center of
  the field of view, where the reflections are concentric about the
  star. The right drawing shows an off-axis star, which in a Schmidt
  telescope produces a reflection that is shifted in towards the
  optical axis. The appearance of the reflections on the detector is
  depicted below the CCD. The drawings are not to
  scale. \label{reflections-drawing}}
\end{figure}

An example of these reflections as seen in the Burrell Schmidt can be
seen in a long observation of Arcturus, shown in Figure
\ref{arcturus-rings}. Along with a completely saturated stellar core
and strong bleeding as expected, we can see scattered light from the
star covering the entire field of view of the detector. The images
also show a bright annulus of light around the star extending out to a
radius of 17 arcminutes. A schematic drawing of the reflections giving
rise to this feature is shown in Figure \ref{reflections-drawing}. In
all of these images the bright ring around the star is caused by
specular reflection of light from the star off the CCD, which then
travels back up to the filter and is reflected again onto the CCD. The
reflected light creates an out-of-focus image due to the additional
path length. This image shows the shape of the telescope pupil, which
is the round aperture of the telescope with a shadow cast by the
Newtonian mirror and its support structure. There are also similar
reflections from other optical surfaces in the telescope, including
the top and bottom of the dewar window and the top surface of the
filter. All of the reflections we observe involve the CCD as one of
the reflecting surfaces since its 10\% average reflectivity is
significantly higher than either the dewar window or the filter.

The observed reflections are not concentric about the star for a
number of reasons. The reflections are shifted radially from the
optical center of the focal plane due to the star's angle of incidence
on the detector. Any inclination of the CCD with respect to the focal
plane, or any curvature to the CCD itself will also cause reflections
to be offset. In the Burrell Schmidt the effect of the star's angle of
incidence is the dominant cause of the offset, and the reflections are
shifted radially inwards toward the optical center. In a Schmidt
telescope the light cone for an off-axis star comes from only a
portion of the primary mirror, since the entrance pupil is defined by
the corrector lens on the telescope and not the mirror. Since the
light from a star comes to the detector from a region on the primary
mirror that is further away from the optical center, the reflected
light will be shifted inwards towards the center of the field. Figure
\ref{reflections-drawing} shows a schematic view the reflection from
an off-axis star, with the light cone first coming to a focus on the
CCD and reflecting first off the CCD and then the filter. The
resulting image is drawn below the CCD and shows the relative position
of the star and the annulus (the drawing is significantly exaggerated
for illustrative purposes).

To characterize the reflections seen on images, we observed a series
of sixteen 450-second exposures of the bright star Arcturus ($M_V =
-0.05$) with the Burrell Schmidt under photometric conditions. For
each of these images the star was positioned in a different portion of
the detector, so that the set of images evenly covered the field of
view. The images were bias subtracted and flat fielded using a sky
flat. The sky flat was generated using over 50 sky pointings
interleaved between science exposures \citep[see,
e.g.,][]{Morrison97,Mihos05}. The images were photometrically
calibrated using stars from the Sloan Digital Sky Survey that were in
the field of view. All data were taken with the Washington M filter,
which was chosen to minimize the contribution of sodium lines from
street lights at 589 nm and the O {\sc I} airglow line at 557 nm
\citep[see][Fig. 1]{Feldmeier02}. The detector used is a $4096 \times
4096$ pixel CCD that was thinned and packaged by the University of
Arizona's Imaging Technology Laboratory (ITL). An anti-reflective
coating was also applied to the detector by ITL. On the telescope the
detector has a pixel scale of $1\farcs 45 \:\mbox{pixel}^{-1}$ for a
field of view of $1\fdg 65 \times 1\fdg 65$.

\begin{figure}
\plotone{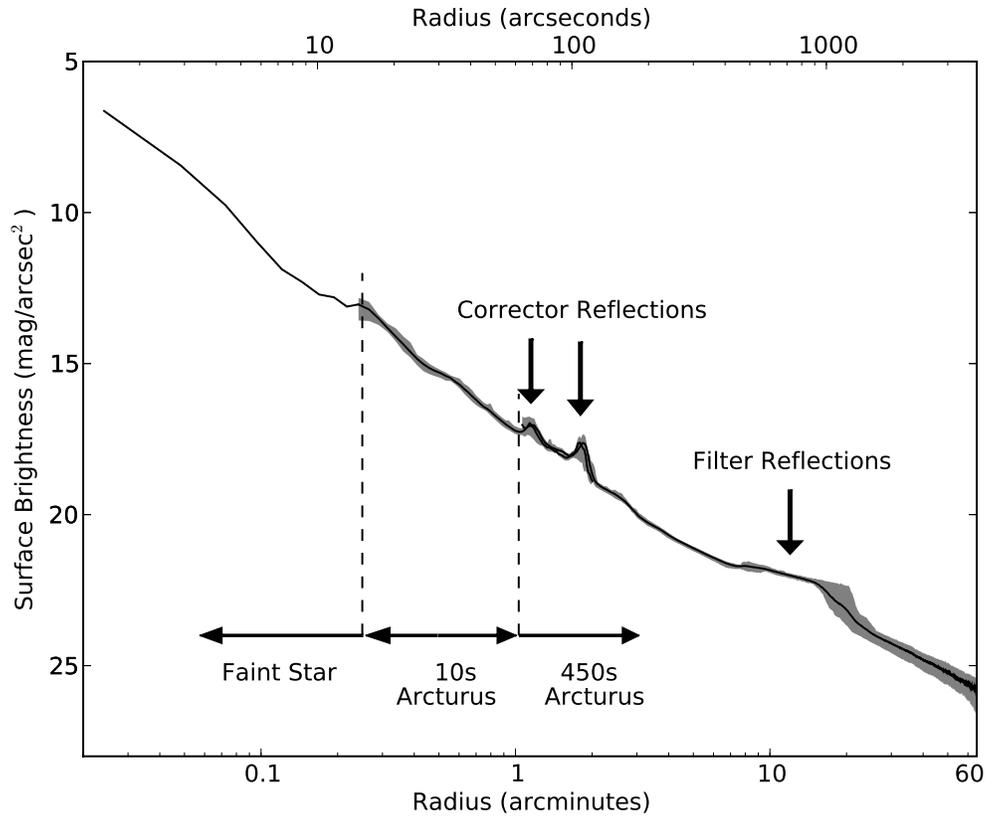}
\caption{Radial profile of Arcturus out to one degree. The profile is
 a composite of the mean of the 450 second Arcturus exposures, the
 mean of the 10 second Arcturus exposures, and a faint background
 star. The dashed vertical lines indicate which observations were used
 to generate the profile in a particlar region. The shading shows the
 variation of the PSF between different positions in the field of view.
 \label{big-psf}}
\end{figure}

A composite radial profile is shown in Figure \ref{big-psf}, using 450
second exposures of Arcturus for the region beyond $\sim 2$
arcminutes, 10 second exposures of Arcturus for the region between 0.2
and 2 arcminutes, and a faint star for the innermost region of
the profile. There are numerous peaks caused by reflections within
various optical elements; the most obvious extend from 0.7 to 2.0
arcminutes and are caused by a reflection within the Schmidt
corrector. This reflection is strongly distorted by the complex
curvature of the corrector and we do not attempt to characterize it.
Bright stars will be masked from the science images out to $\sim 2$
arcminutes in order to eliminate all of these features.  The corrector
is also responsible for the ``Schmidt ghost,'' which is a diffuse image that
appears reflected across the optical axis from bright stars. The ghost
is the result of light from a star reflecting off the CCD, travelling
back up the telescope, reflecting off the corrector, and ultimately
returning to the CCD. This ghost was masked out of all calibration and
science images.  The most distinct reflections beyond 2 arcminutes are
are caused by the dewar window and the filter, and appear as bumps in the PSF
at roughly 2.5 and 10 arcminutes from the star. These are the
reflections that will be modeled and removed.

\begin{figure}
\plottwo{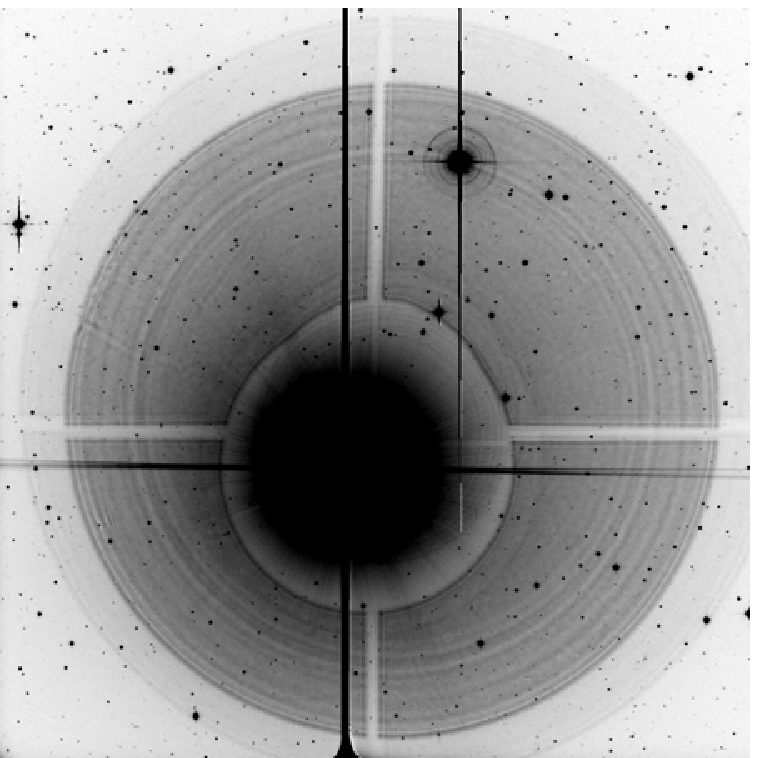}{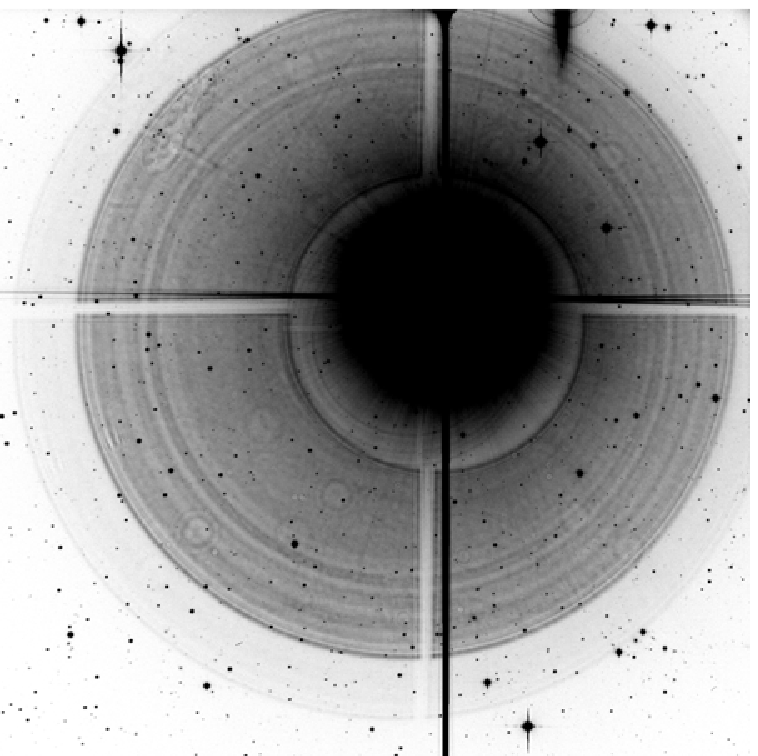}
\caption{Two 450-second exposures of Arcturus, with the star
  positioned in opposite corners of the field of view. The optical
  center is towards the top right of the image on the left, and
  towards the bottom left on the right image. The images saturate to
  black at $\mu_V = 21.2$. \label{arcturus-nosub}}
\end{figure}

The sources of the reflections can be confirmed by using the size of
the reflections to determine the extra path length traveled by the
reflected light. The size of the two largest reflections correspond to
a reflecting surface 3.6 cm away from the CCD, which matches the
position of the filter. The bottom surface of the filter (the side
closest to the CCD) causes the brightest reflections, and can be
easily seen in all the example images. The top surface of the filter
produces a much fainter reflection, which is barely apparent (see
Figure \ref{arcturus-nosub}) as a ring slightly beyond the bright
annulus of the bottom surface's reflection. There is also a reflection
off the bottom surface dewar window, which lies between the CCD and
the filter, and a fainter reflection off the top of the dewar window.
The size of the bottom reflection indicates that the window is 0.6 cm
away from the CCD as expected. The reflection off the top of the dewar
window was faint enough to be far below the noise level on realistic
foreground stars near the science fields, and so we do not attempt to
model it. Similarly multiple bounces between the reflecting surfaces
and the CCD will contribute a small amount of light at large radii,
but using the measured reflectivities we calculate that all of the
secondary reflections will have surface brightnesses fainter than
$\mu_V = 30$ mag/$\mbox{arcsec}^2$ in our Arcturus images, well below
our detection limit. A summary of the reflecting surfaces and the
brightness of their reflections is presented in Table
\ref{param-table}.

These reflections have been modeled in Zemax, an optical ray-tracing
program, to confirm their source and the causes of the offset between
stars and their reflections. The model of the telescope included all
optical surfaces in the light path, including the Schmidt corrector
plate, primary and Newtonian mirrors, filter, dewar window, and
CCD. In the Burrell Schmidt there is also a contribution to the shift
from the convex curvature of the detector, which is on the order of a
$100\:\micron$ height difference between the center and the edges of
the CCD. The measurement of the curvature was obtained from focus
images, where we found the focus position that produced the smallest
star profiles in each corner and in the center of the chip. Using this
information we were able to calculate the expected centers of the
reflections, and therefore their offset from the star. The ray-tracing
found reflection centers that matched the observed positions to within
ten pixels, without accounting for any inclination of the filter or
CCD.

Using observations of Arcturus to measure the offset between the
center of the star and the center of its reflection we determined that
the offset could be modeled as a linear function of position on the
detector. For instance, a star on the optical axis has concentric
reflections, and as one moves away from the axis the distance between
the star and the center of its reflection increases
linearly. Combining fits of the offset as a function of row and of
column on the detector can account for both radial shifts (caused by
the angle of incidence) and linear shifts (caused by tilts of the
optics or detector).

\begin{figure}
\plotone{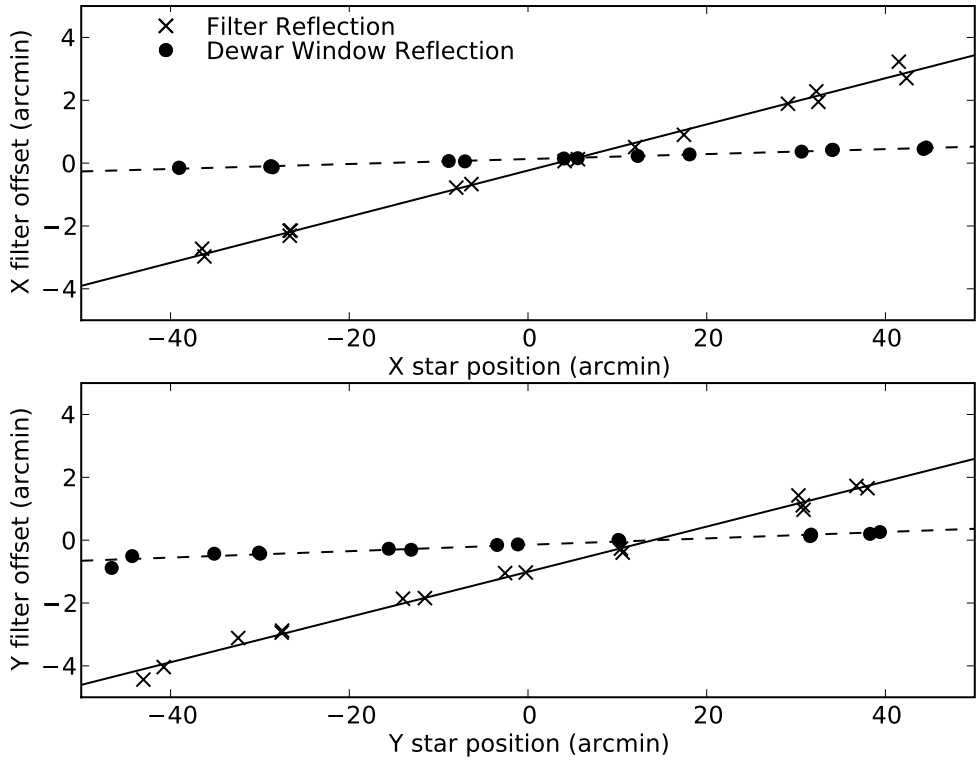}
\caption{Offsets of the reflections from the star in x and y
  coordinates over the field of view. The solid and dashed lines are
  linear fits.\label{offsets-fig}}
\end{figure}

The measured offsets of the reflections are plotted in Figure
\ref{offsets-fig}. The bottom of the filter and the dewar window
reflections are both shown. This confirms that the shifts are a linear
function of position, and that the curvature of the CCD has not added
a significant nonlinear component as one would expect from an
extremely concave or convex detector. The figure also shows that the
offsets can be measured empirically simply using bright-star
observations that cover the field of view. Although ray-tracing was
helpful in confirming our understanding of the behavior of the
telescope, it is not necessary for modeling the reflections.

To determine the brightness of the reflections, we first roughly
measured the difference in brightness between regions just inside and
just outside the edge of the reflection with IRAF's imexam tool. We
then recalculated the radial profile after subtracting off the
reflection to see if any evidence of the reflection could still be
seen. This first estimate was further refined by making small changes
to the fraction of light reflected that our model specified, then
repeating the process of applying the reflection removal and
calculating the resulting radial profile. We continued to adjust the
reflection in this way until the profile was featureless around the
area of the reflection we were removing. This process was performed
for a single image, then the results were applied to all of the other
observations using the same value of the fraction of reflected
light. The resulting profiles in the other images also showed no
features in the reflection region, which indicates that our
reflectivity measurement is not biased by only using one image in the
process. This also indicates that the reflectivity of the various
optical surfaces is position-independent to within our errors.

To summarize our parameters, each reflection is modeled by an offset
from the star, which is a linear function of both row and column on
the detector, a constant circular size, and a brightness, which is the
fraction of the total intensity of the star that reaches each pixel in
the reflection area.

We determined these parameters for three surfaces in the Burrell
Schmidt: the top and bottom surfaces of the filter and the bottom
surface of the dewar window. All of the reflections also have an inner
shadow in the center caused by the obstruction of the pupil by the
Newtonian mirror. The position of the shadow can be modeled in the
same way as the other reflections. The reflection from the bottom of
the filter shows this shadow most clearly. However, the shadows from
the filter top and dewar window bottom reflections are not visible in
our Arcturus observations since they are both very faint relative to
the profile of the star at their respective radii. The inner edge of
the dewar window reflection is close enough to the star that it will
be masked, and the oversubtraction of the extremely faint filter-top
reflection will be negligible in the inner regions of the profile.

\begin{figure}
\plottwo{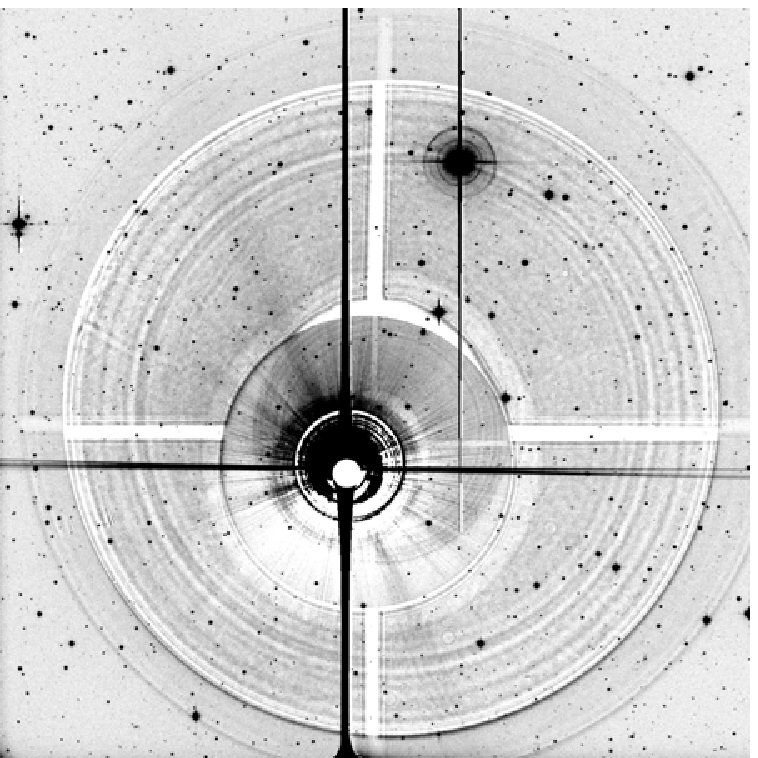}{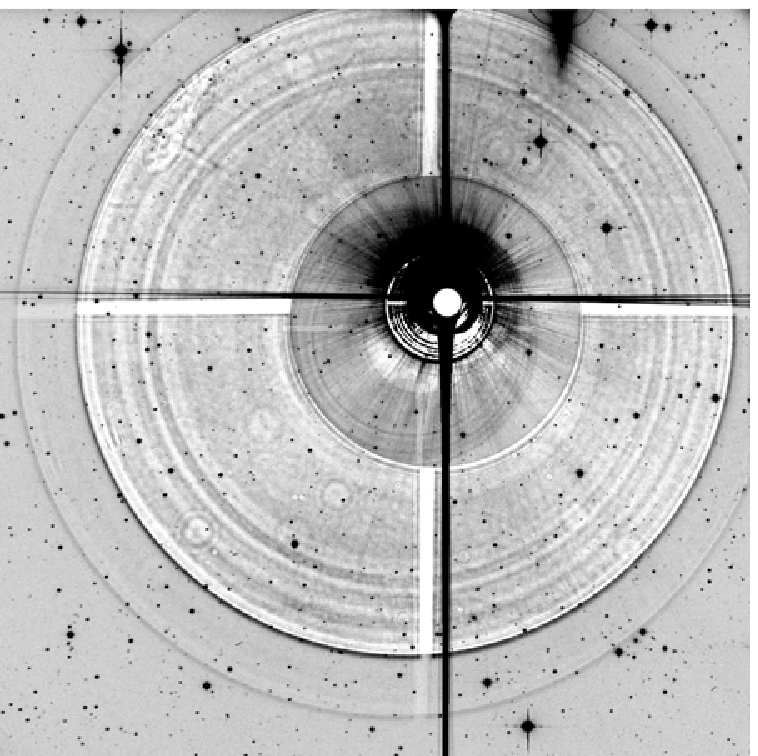}
\caption{Same images as in Figure \ref{arcturus-nosub} after removal
  of the reflections and PSF. The images saturate at $\mu_V = 22.1$,
  roughly half the brightness of the images in Figure
  \ref{arcturus-nosub}. \label{arcturus-sub}}
\end{figure}

Figure \ref{arcturus-sub} shows the result of removing the reflections
from the images of Arcturus shown in Figure
\ref{arcturus-nosub}. The reflections are completely removed except
for some small amplitude rings and the shadows caused by the spider
supporting the Newtonian mirror. These faint structures will be
smoothed out by coadding multiple dithered images, since in each image
the structures will have moved slightly.

The radial profile before and after removing the reflections can be
seen in Figure \ref{psf-fig}. The upper line shows a mean of the
uncorrected radial profile of the sixteen bright star images, while
the middle dotted line shows the mean of profiles after removing the
reflections from the individual images. The shaded area shows the
scatter between the images, and extends between the maximum and
minimum intensity of the sixteen profiles at a given radius. The
profiles are photometrically calibrated with other stars in the field
and are not affected by the saturation of the core of Arcturus. The
regions of the profile corresponding to removed reflections are
indicated by the horizontal lines at the top of the Figure
\ref{psf-fig}. These are not necessarily the exact positions of the
removed features, since the reflections are offset from the center of
the star and hence are not radially symmetric about the star. This
also causes the radial profile to show a smooth transition instead of
the sharp cut-off at the reflection edges as seen in the images.

\section{Atmospheric and Internal Scattering}
\label{scattering}

\begin{figure}
\plotone{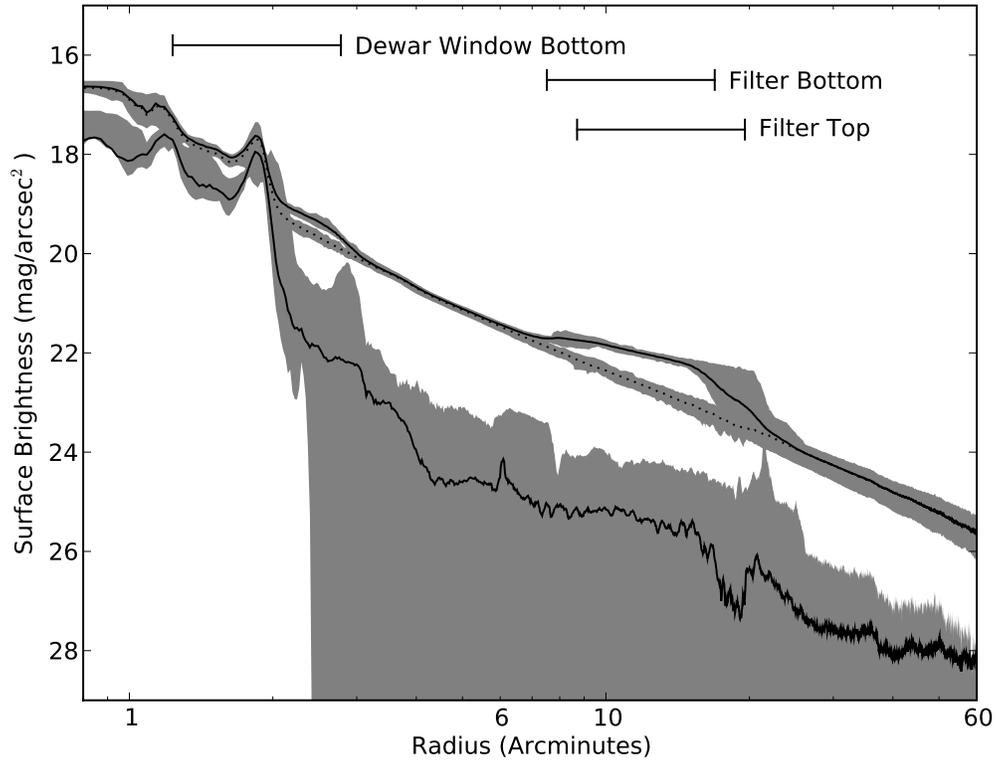}
\caption{The radial profile of Arcturus in the original images (top
  solid line), after removing reflections (middle dotted line), and
  after removing both reflections and scattered light (bottom solid
  line). In the top two profiles the line indicates the mean of the
  profiles of individual images, while the bottom line is the profile
  of coadded image. The shaded regions indicate the maximum scatter
  between individual observations. The horizontal lines indicate the
  sizes of the reflections, although the exact area affected will vary
  slightly since the reflections are not concentric about the
  star. The read noise level for single exposures is equivalent to
  27.6 $\mbox{mag}\,\mbox{arcsec}^{-2}$.
  \label{psf-fig}}
\end{figure}

After removing the modeled reflections from each pointing, our bright
star images were coadded to obtain a mean point-spread function for
all positions on the detector, which is shown by the middle line in
Figure \ref{psf-fig}. We expect that point-spread function after
reflection removal to be independent of position in the field of
view. This is attested by the small scatter between the profiles of
the different pointings, as shown by the light gray shading.

The PSF can be divided at 5 arcminutes into to two regions with
significantly different behavior. Inside of 5 arcminutes the behavior
of the profile is determined by the effects of diffraction, turbulent
scattering in the atmosphere, and complex internal reflections in the
corrector. In this region the profile exhibits a steep dependence on
distance from the star that becomes shallower further
out. Approximating different parts of the profile inside of 5
arcminutes as power laws produces varying results, with power law
slopes ranging from -3 to -2.4.

Inside of 10 arcseconds the profile should be dominated by the
behavior of turbulent scattering in the atmosphere, along with
diffraction. Assuming the turbulence follows Kolmogorov statistics
this inner PSF should be fit by a Moffat function
\citep{Racine96}. However, in our data the $1\farcs45$ pixels do not
adequately sample the PSF in this inner region and we can only
observe the outer wings of the turbulent scattering.

Between roughly 10 and 200 arcseconds a number of bright reflections
appear in the profile. These reflections are caused by multiple
bounces within the corrector lens, and exhibit complex and varying
shapes due to the figuring of the corrector. This part of the radial
profile has been measured in many other telescopes with results often
indicating an $r^{-2}$ behavior \citep{King71,Surma90,Racine96} or
slightly steeper (\citealt{Gonzalez05}; and see \citealt{Bernstein07}
for a more complete review). Our measurements find a steeper behavior
than $r^{-2}$ and closer to $r^{-2.4}$. The complex reflections make
it difficult to determine the source of this steep profile. We do note
that diffraction would show a steeper profile, since the
light from diffraction by an annular aperture falls off as
$r^{-3}$. Despite this rapid fall-off, the light from diffraction can
still be significant at radii up to an arcminute or more. Note that
the sharp diffraction spikes caused by the Newtonian spider structure
are not the issue here; those spikes are masked out of the radial profile.

Beyond 5 arcminutes the PSF begins to fall off less rapidly, and can
be fit roughly with a power law of index -1.6. Few measurements of the
PSF on other telescopes extend out to these radii. \citet{King71}
finds a power law index of -2 at these radii. King refers to the
$r^{-2}$ behavior as the aureole. There is no consensus as to the
source of the aureole, though there has been speculation about
scattering due to dust, microripples in the primary mirror, and
multiple-bounce diffuse reflections \citep{Racine96}.

Similarly we cannot provide any confident explanation of the
$r^{-1.6}$ behavior we observe.  One hypothesis could be that improper
background subtraction lead to a bias in the profile, but this cannot
be supported. While the behavior of the PSF beyond $\sim 20$
arcminutes is sensitive to improper background subtraction, the region
inside of 20 arcminutes should be dominated by scattered light from
the star. At 10 arcminutes from the star, our 5 ADU uncertainty in the
background subtraction will vary the profile by less than 0.01
magnitudes, while at 40 arcminutes the uncertainty in the background
subtraction contributes to an uncertainty of 0.1 magnitudes. It is
therefore unlikely that the $r^{-1.6}$ behavior could be an artifact
of the background subtraction.

Regardless of the source of the scattered light, we model the point
spread function with a third-order polynomial in log space, fit to the
mean PSF between 2.4 and 60 arcminutes. This region contains $1.2\%$
of the total light from the star, which is roughly consistent with
measurements made at other telescopes \citep{Bernstein07}. Beyond
one degree the amount of light from Arcturus becomes difficult to
distinguish from the sky background. Since no science fields are close
enough to a star as bright as Arcturus to cause significant
contamination at these radii, we are not concerned with modeling
beyond one degree.

The result of subtracting the PSF can be seen as the bottom line of
Figure \ref{psf-fig}. The line indicates the profile after coadding
all of the filter- and PSF-subtracted Arcturus observations, while
again the gray region delimits the spread of the profile from
individual images. This coadd approximates the behavior one would see
in images produced by coadding dithered observations, which will
smooth out any remaining high spatial frequency features in the
stellar profile. 

In modeling the PSF we chose to be somewhat cautious about avoiding
oversubtraction. At any given radius the observed profile of a single
star will have some scatter which will be asymmetric around the mode,
in the sense that there are more likely to be features with positive
intensities (e.g. nearby stars in the field or the star's faint
diffraction spikes) or than features with negative intensities
(e.g. read noise, which is symmetric). Because of this our fit was
designed to err on the side of undersubtracting, and as a result the
subtracted PSF in Figure \ref{psf-fig} still shows some positive
features such as diffraction spikes inside of 5 arcminutes.

\section{Science Reduction}
\label{sciencereduction}

In order to assess the systematic biases and artifacts introduced by
different methods of star subtraction we have reduced deep
observations of the Virgo cluster with both our PSF modeling technique
and other more common methods, including using no star subtraction and
subtracting a static PSF. In many situations, including point source
photometry and localized surface photometry, the scattered light
contribution from nearby stars is often ignored and treated as a
generic background. This is reasonable in these situations because the
variability in the PSF as a result of reflections only affects a very
small fraction of the total light from the star. Nearby stars are also
often ignored when measuring the total integrated magnitude of a
galaxy, for much the same reason. In faint extended surface
photometry, such as isophote fitting of galaxies, star subtraction is
usually performed with a PSF that is static across the field of
view. This can be a reasonable assumption for telescopes with small
fields-of-view in which the PSF does not vary much, but with very
large detectors modeling of the variable PSF becomes much more
important. 

To compare the effectiveness of our modeling to other reduction
strategies we have performed a series of reductions on deep
observations of the Virgo cluster. We observed a region near the
center of the Virgo cluster for eight nights under photometric
conditions and no moon, obtaining 52 science exposures of 900 seconds
each in the Washington M filter. The images were bias subtracted,
flattened with the same flat as had been used for the Arcturus data,
and the magnitudes were transformed to Johnson V magnitudes. We then
developed a catalog of stars in and near the target field from the
Tycho 2 and SDSS databases. For every image we selected all the stars
from these catalogs within 72 arcminutes of the center of the image
and constructed a model PSF for each star, scaled to the star's total
flux. For stars that fell onto the detector, we also constructed a
model image of the reflections generated by each star. We coadded all
of the model PSFs and reflections to obtain an image of the total
scattered light in a field, and we subtracted this image from the
science observation.

After removing the scattered light, we sky-subtracted the images using
the method described by \citet{Mihos05}. In this technique the images
were registered, then a sky image was constructed from each one by
masking out all objects. Planes were fit to the individual sky images
and an iterative procedure adjusted the planes in order to minimize
the discrepancy between exposures after removal of the sky planes.

\begin{figure}
\plottwo{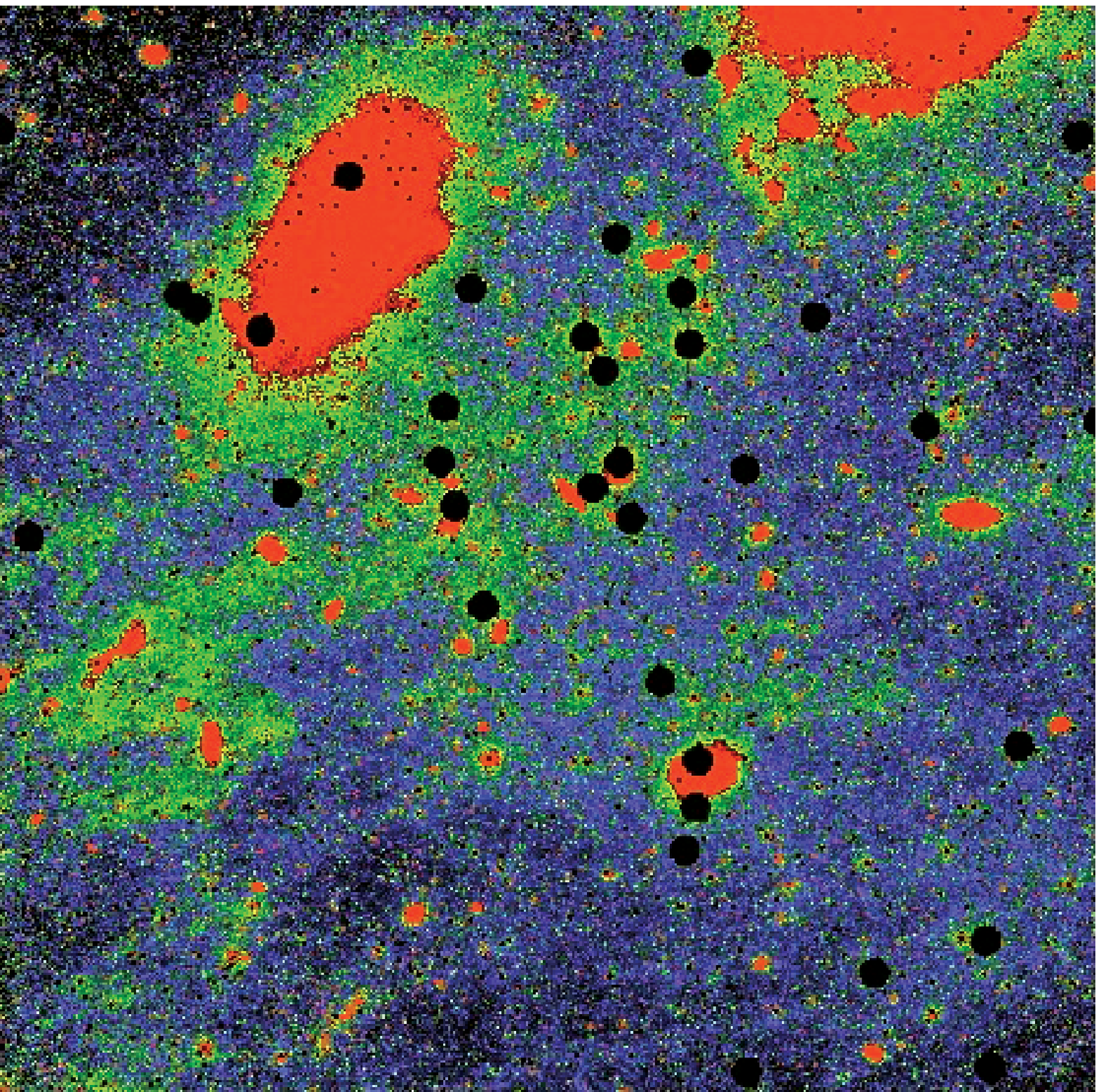}{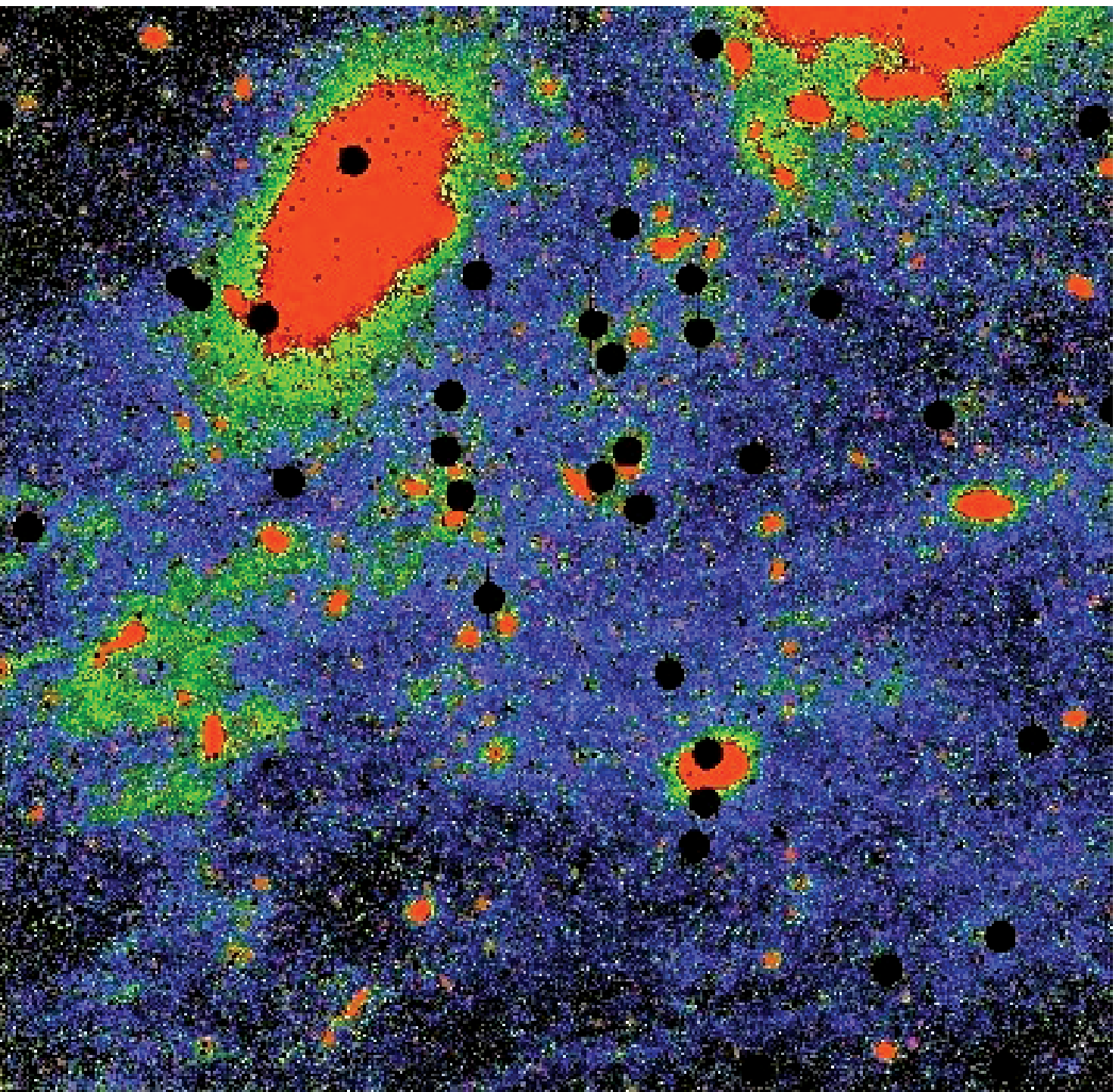}
\caption{Virgo mosaic without any star subtraction is on the left, and
  on the right is the mosaic with full reflection modeling and star
  subtraction.  North is up and east is to the left, and M87 is in the
  upper left. The mosaic is $\sim 2.5 \degr$ on each side. In both
  images red is $\mu_V < 27.1 \,\mbox{mag}\,\mbox{arcsec}^{-2}$, Green is
  between 27.8 and 28.0, and blue is fainter than 28.0 
  $\mbox{mag}\,\mbox{arcsec}^{-2}$
  \label{goodred}}
\end{figure}

The resulting mosaic of the Virgo cluster is shown on the right side
of Figure \ref{goodred}. Much of the eastern region of the mosaic is
contaminated by high-latitude galactic dust, which reflects light from
the Galactic disc \citep[e.g.,][]{Sandage76,Witt08}. It is unlikely that any of
the green-colored structure in this area is intracluster light. Faint
ICL structures can indeed be seen emanating from M87, as was
previously observed in \citet{Mihos05}.

\begin{figure}
\plotone{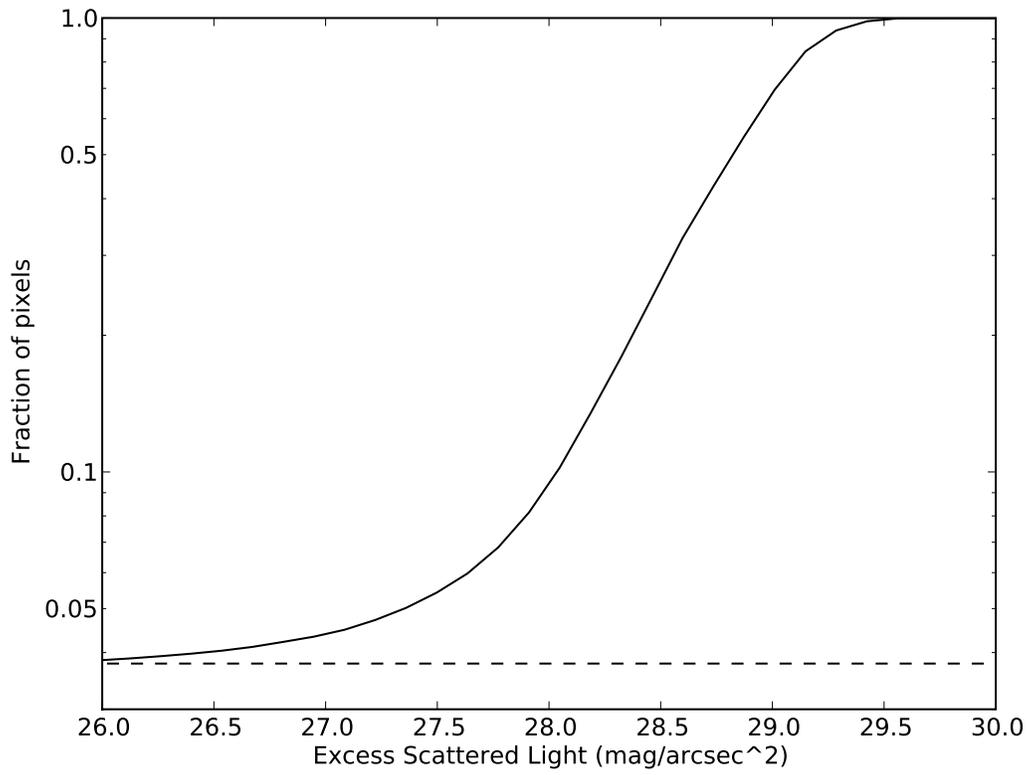}
\caption{Cumulative histogram of the scattered light observed from
  stars in the mosaic of Virgo. The solid line indicates the fraction
  of pixels that received scattered light at a given surface
  brightness. The dashed line shows the fraction of pixels that are
  masked due to contamination by bright stars. \label{histogram}}
\end{figure}

\begin{figure}
\plotone{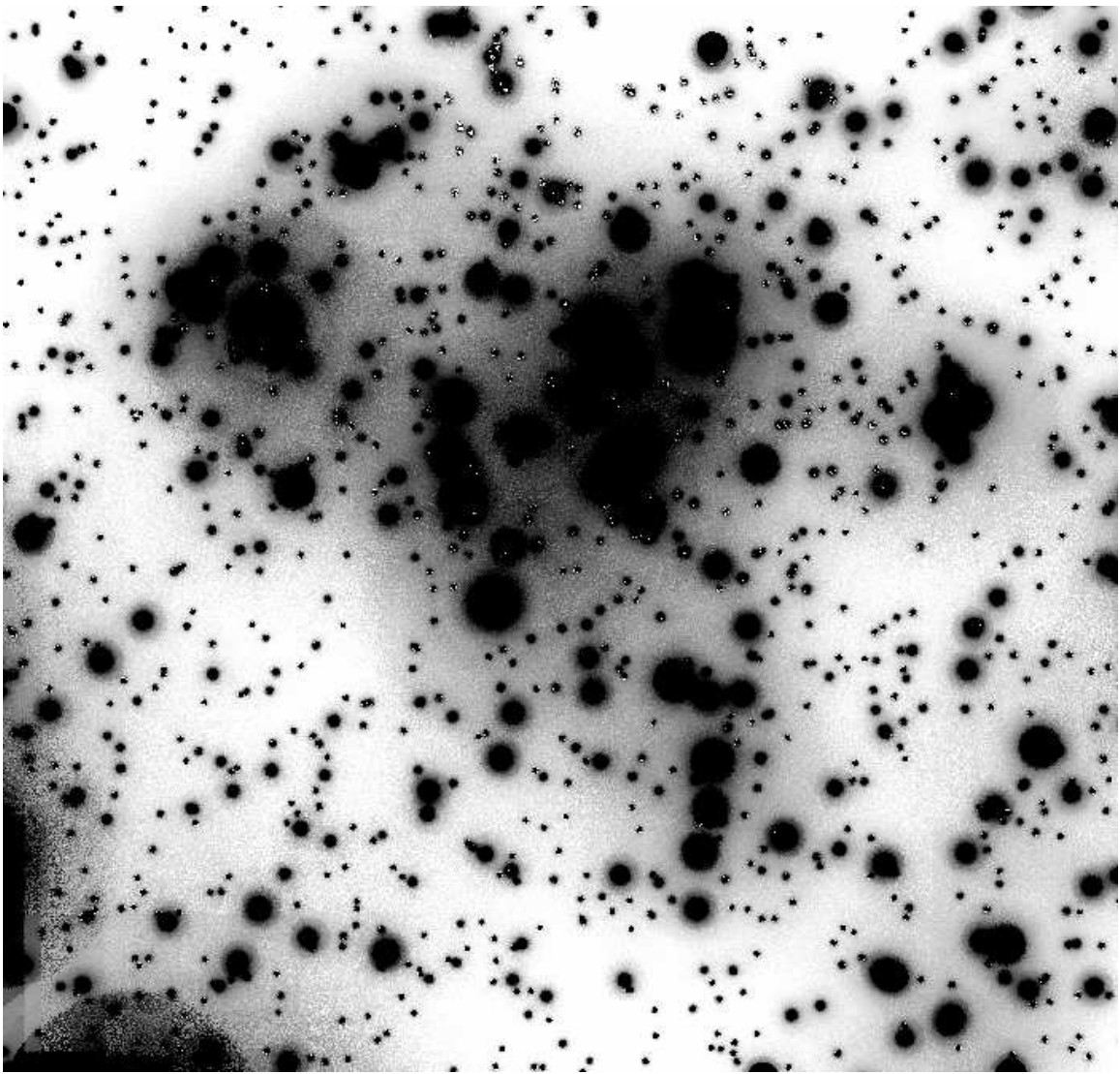}
\caption{This image is the difference between the reduction with no
  star subtraction and the image with the fully modeled reflection and
  PSF subtraction. Black indicates an excess of light in the image
  without star subtraction, and saturates at 2.5 ADU ($\mu_V =
  27.8$). \label{diff2}}
\end{figure}

To emphasize the impact of the scattered star light we performed one
reduction of the data without any reflection removal or star
subtraction. This is shown on the left in Figure \ref{goodred}. The
difference between the properly star-subtracted image and the
reduction without star subtraction is shown in Figure \ref{diff2}. The
most significant contribution to the excess light comes from three
eighth-magnitude stars to the west of M87. In total, the flux from
foreground stars in the field is equivalent to a single star of
magnitude $M_V = 3.7$. As discussed in \S\ref{scattering}, since at
least $1.2\%$ of the star light is scattered to radii beyond
2.4$\arcmin$, there is significant diffuse light across the entire
field. Figure \ref{histogram} shows the fraction of the mosaic
contaminated by scattered light at a given level. This shows that
50\% of the image has scattered light at the $\mu_V = 28.8$ level, and
10\% receives more than $\mu_V = 28.0$ of scattered light.

This excess light not only limits depth to which we can detect faint
structure, but also strongly biases the background
subtraction. Because we fit planes to individual images, the
plane-fitting will attempt to remove the excess light from stars in
the mosaic center. This distorts the background planes and
significantly degrades the accuracy of the background subtraction in
the outer parts of the mosaic. This has the potential to distort
isophotes in the outer regions of galaxies where proper background
subtraction is crucial. In the images shown we subtract the same
background planes from both reductions in order to ease comparison
between the two. Even with the proper background subtraction the outer
regions of M87 show considerable variation between the two reductions.
Toward the southern end of M87 the image without star subtraction
shows the galaxy extending much further than the star-subtracted
image. Any attempts to measure the galaxy's profile at these distances
must therefore carefully account for scattered light effects.

The improvement in the final mosaic made by modeling the reflected
light is perhaps not as dramatic in the Burrell Schmidt as it would be
in telescopes that are less optimized for deep surface photometry. The
anti-reflective coatings on the optics in the Burrell Schmidt
significantly reduce the reflected light compared to other optical
telescopes, including many large multi-user facilities. The
reflectivities of the optics in the Burrell Schmidt are all less than
$0.8\%$, while reflectivities around $2$ - $4\%$ are much more
common. We have attempted to emulate the images that would be produced
by a telescope with higher-reflectivity optical components by using
our model to generate the additional light that would be reflected. We
add this to the science observations to obtain an image that is more
representative of what would be observed on an unoptimized
telescope. Without any star subtraction the image with bright
reflections shows an excess of light of almost 7 ADU around regions
with bright stars, and on average an excess of 4 ADU spread diffusely
across the mosaic. This diffuse component is equivalent to a surface
brightness of $\mu_V = 28.0 \,\mbox{mag}\,\mbox{arcsec}^{-2}$.

Although the reductions without star subtraction make the most
dramatic comparison, it is not representative of common practice in
deep surface photometry. The most widely used method of treating
foreground stars is to subtract a PSF that is static across the field
of view. We investigate the biases and artifacts caused by this
technique by using our images with the exaggerated reflections and
subtracting a mean, static PSF from each star.

\begin{figure}
\epsscale{0.6}
\plotone{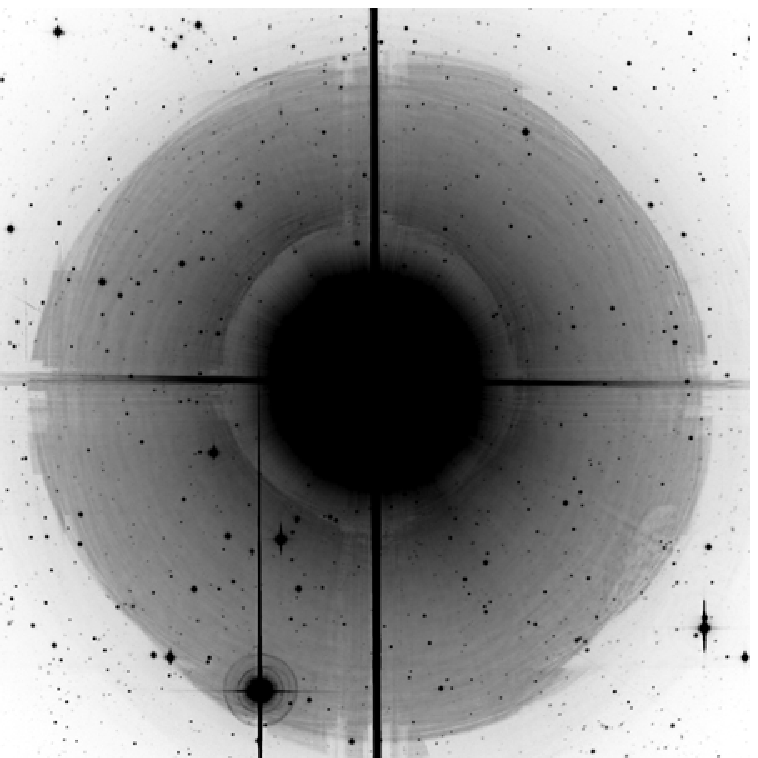}
\epsscale{1.0}
\plottwo{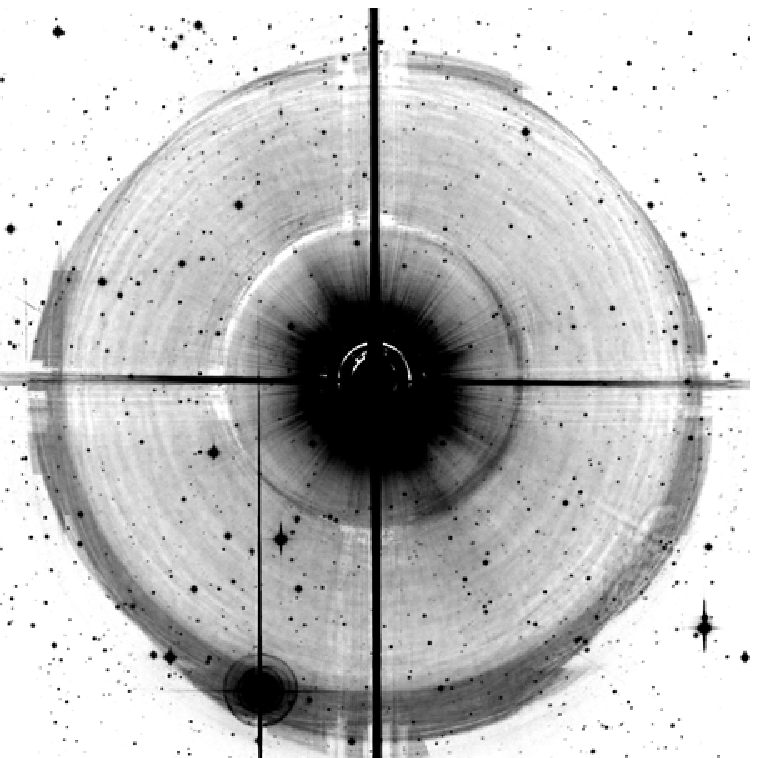}{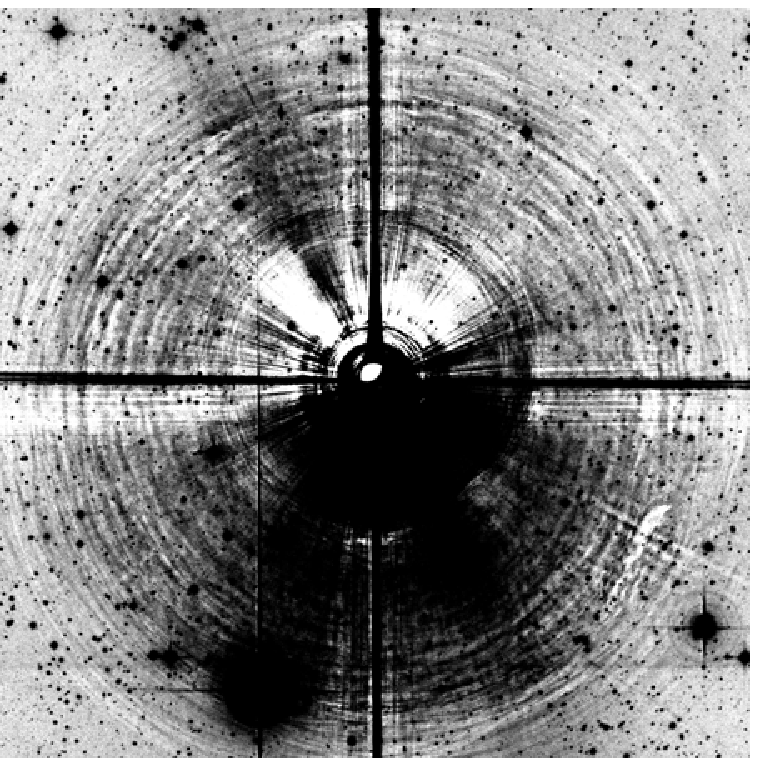}
\caption{Top: Coadded mosaic of all 16 Arcturus pointings without any
  star subtraction. Left: Coadded mosaic after removing an average
  radial profile generated from a single Arcturus image. Right:
  Arcturus mosaic after subtracting the position-dependent PSF from
  each individual pointing. The top image saturates to black at 900
  ADU above the background ($\mu_V = 21.5$), the left image saturates
  at 400 ADU ($\mu_V = 22.3$), and the right image saturates at 50 ADU
  ($\mu_V = 24.6$).
\label{arcturus-comparison}}
\end{figure}

\begin{figure}
\plottwo{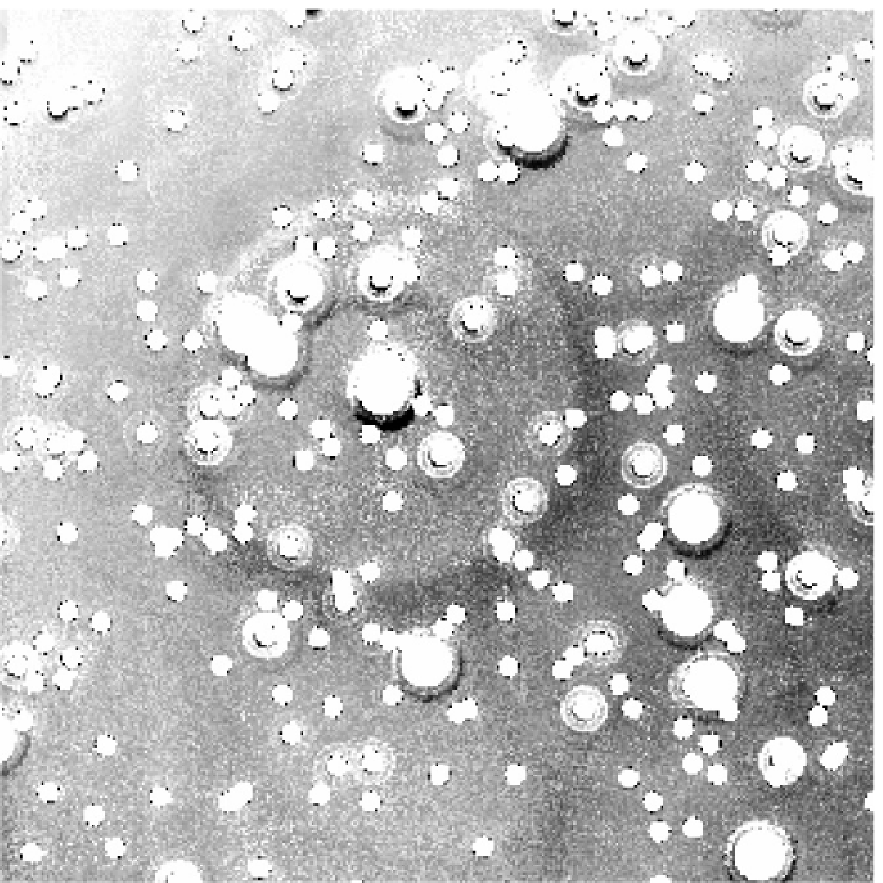}{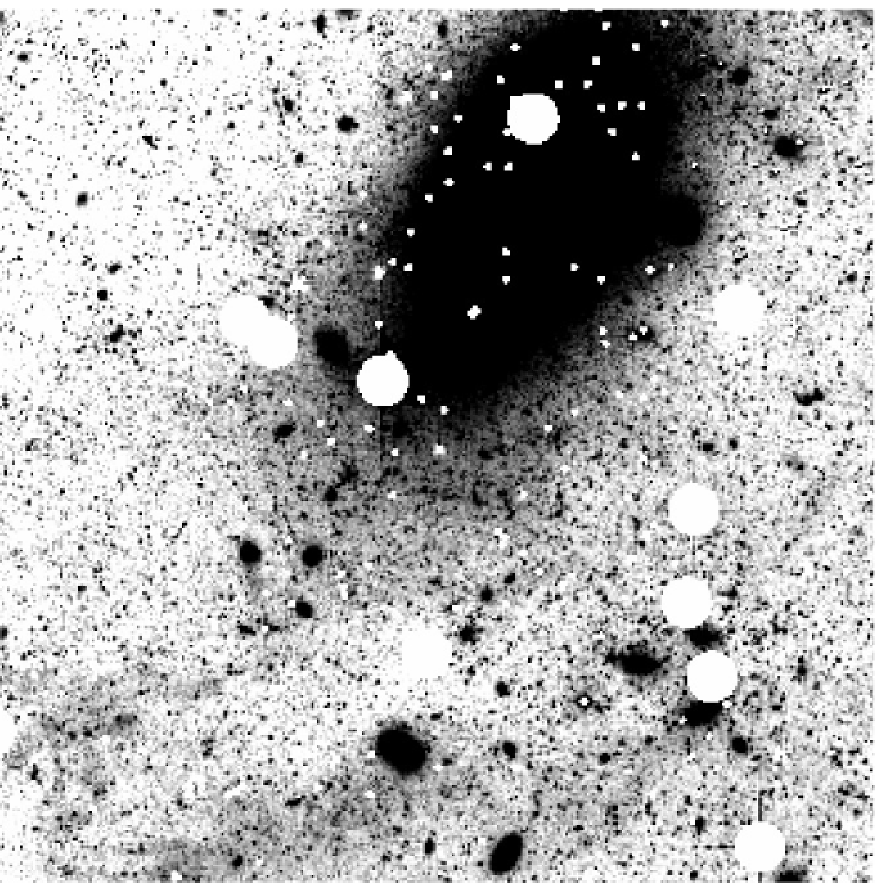}
\caption{On the left is the difference between the image with
  exaggerated reflections that had a static PSF removed and the image
  with the full reflection modeling. Black in the image indicates an
  excess of light in the reduction with the static profile. The image
  saturates to black at 1.5 ADU ($\mu_V = 29.0$). On the right is the
  same region (showing M87) in the image with full reflection
  modeling.
  \label{static-zoom}}
\end{figure}

This static profile creates significant errors near the edges of the
reflections around each star, and has the potential to mimic
sharply-defined tidal structures in adjacent galaxies. A comparison
between different PSF subtraction methods is shown in Figure
\ref{arcturus-comparison}, which uses sixteen dithered exposures of
Arcturus. In the top panel a normal coadd is shown with no star
subtraction. The left panel shows the coadd after subtracting a static
profile from each image, and the right shows Arcturus after
subtracting modeled reflections from each exposure. The bright ring in
the left image is illustrative of the artifacts that using a static
PSF can introduce. The ring has the same brightness as the filter
reflection. The artifacts of static-PSF star subtraction can also be
seen in a mosaic of the Virgo observations. Figure \ref{static-zoom}
shows a region around M87 where a bright star has caused a ring which,
though faint at the $\sim 1$ ADU level, could easily mimic a tidal
feature. While such spurious features might be recognized by their
proximity to a bright star, more complex structures due to the
overlapping wings of a number of stars might be more difficult to
clearly identify as artifacts.

One concern that can be raised about our method is that we only
correct for scattered light from bright stars in the field, while
bright galaxies will also contribute scattered light. This
contribution can be significant since, for example, the integrated V
magnitude of M87 is 8.63 \citep{RC3}, which is comparable to some of
the brighter stars in our field. The most technically correct method
of accounting for this scattered light would be to deconvolve the
images with our spatially varying model for the reflections and the
PSF, but this is an extremely computationally-intensive operation.
Instead, we simply estimate the severity of the effect by calculating
the scattered light recieved at a given pixel as if the original image
was a true measure of the light distribution in the field (unaffected
by scattering).

We can perform this estimation using M87 as an example. At the a
radius of $10\arcmin$ from the center of M87, similar to the distance
to the NGC 4476/4478 to the west, the scattered light has a surface
brightness of roughly $\mu_V = 26.7$. Subtracting this excess light
from the measured surface brightness of $\mu_V = 25.8$ at $10\arcmin$
along the minor axis results in a change of roughly 0.5
magnitudes. Further out at $15\arcmin$ the scattered light falls to
$\mu_V = 28.5$, which would lead to a 0.2 magnitude correction in the
measured surface brightness at that distance along the minor axis. At
even larger radii from the galaxy the scattered light quickly falls
below the noise threshold. Thus in our observations of the ICL we can
safely ignore the scattered light from galaxies, but observing
programs that focus on the light profiles of galaxies themselves at
intermediate radii may require a more careful treatment
\citep[e.g.][]{deJong08}.

\section{Summary and Application}
\label{summary}

We have presented a method for characterizing and removing reflections
and scattered light in the post-processing of images. We found that
reflections off of the CCD produce annular rings in the images after
they reflect back off of either the filter or dewar window. The rings
are circular and are uniform in their brightness to roughly 5\%, and
are not necessarily axisymmetric about the central star. However, the
offset between the star and the center of the rings is a linear
function of the star's position on the detector. Using this
information we build a model for the reflections and the diffuse
scattered light from each star and use it to remove this excess light
from our observations of the Virgo cluster.

By comparing our reduction techniques to other methods we have shown
that star subtraction is critical for faint surface
photometry. Omitting star subtraction from the reduction, even on an
optimized telescope, will create significant errors in background
subtraction and will distort the appearance of the outer regions of
galaxies. We have also shown that performing star subtraction with a
PSF that does not vary over the field of view can be a reasonable
reduction method, but runs the risk of creating artifacts that look
conspicuously like tidal structures. Modeling the changing PSF across
the field is the safest way to ensure that these artifacts do not
appear.

Porting this method to other telescopes requires three main sets of
calibration observations: a set of bright star observations across the
field of view to measure the behavior of the reflections in an
instrument, one or more deep observations of a bright star to measure
the smooth component of the extended PSF, and a small number of
periodic checks during an observing run to confirm the stability of
the reflections.

The first step in developing the reflection model is to obtain 10-16
bright star observations to ensure that the reflections in a
particular instrument and telescope are linearly related to the
position of the central star. These exposures only need to be long
enough to make the reflections visible. Ray tracing could be used to
confirm the linear behavior of the reflections, but in most
instruments it cannot substitute for the empirical measurements of the
reflection offsets. To obtain the required accuracy from ray tracing,
assuming spacings of a few centimeters between the CCD and the
reflecting surfaces, the surfaces would need to be aligned to within
one to two arcminutes of the optical axis. For a four-inch filter this
would mean that the filter could only deviate 100 $\mu m$ from
parallel. In instruments where these tolerances are not met the only
reliable way to measure the reflection locations is with bright star
observations.

Once it is known that the reflections behave linearly in an
instrument, the locations of the reflections only need to be checked
periodically throughout an observing run with a small number of bright
star observations. These observations also do not need to be very
long, but should be performed at least once per observing run, if not
more frequently. 

Longer observations are needed to measure the smooth component of the
extended stellar PSF, but fewer exposures are necessary since this
component of the PSF should not vary across the field of view. These
observations can be performed much less frequently than the reflection
checks, and it should only be necessary to repeat them when
significant changes to the optics are made, such as realuminization of
the mirrors or changes of the filters or optics.

It is also important to underscore the importance of dithering
observations of a field when attempting to perform surface
photometry. Throughout this work we have demonstrated the complex
structure of the instrumental PSF (speaking broadly to include the
reflections) and the variation it exhibits across the field of
view. Though we can successfully remove the brightest components of
this structure, we still rely on dithering to average out smaller
scale features such as the shadow of the Newtonian spider and the
rings caused by the corrector. For this to be effective the
movement between exposures needs to be on the order of the size of the
features to be removed. This ensures that the features move relative
to the star enough to average out instead of adding
constructively. Dithers of only a few pixels, as are commonly used for
the drizzle routine, will likely be ineffective at reducing reflection
artifacts.

It is worth reiterating the simplicity of the assumptions we have
made. Two physical insights are at the center of the process:
reflections are out-of-focus images, and out-of-focus images exhibit
the illumination pattern of the telescope aperture. The curvature and
inclination of the various reflecting surfaces only cause small
perturbations to this image, resulting primarily in spatial shifts. We
have found that linear modeling is sufficient for removing these
artifacts from simple direct imaging applications, and that the
resulting improvements in low surface brightness imaging is worth the
extra observing and data reduction effort.

\acknowledgments

This research was supported by the NSF through grants ASTR 06-07526
and ASTR 07-07793 to J.C.M.

\appendix

\section{Hardware Upgrades to the Burrell Schmidt}

Though in this work we are primarily concerned with improving the
sensitivity of the Burrell Schmidt with software, numerous upgrades
to the telescope hardware have also played a significant role in the
optimization of the telescope for deep surface photometry. 


One of the telescope's main strengths is its simple optical
layout. Since the camera is a direct-imager, there are only three
transmissive optical elements in the light path: the corrector lens at
the top of the tube, the filter, and the dewar window. The dewar
window also serves as a field flattener, which alleviates the need for
another lens. Compared to the number of components needed for a
reimaging camera, this setup significantly reduces the number of
reflections that would need to be modeled and removed. The filter and
dewar windows also have antireflective coatings which further reduce
the amount of scattered light. The filter was produced and coated by
Barr Associates, with a specification for a reflectivity less than
$0.5\%$.

The telescope's closed tube significantly reduces the amount of stray
light that reaches the detector. In an open-tube Cassegrain design,
light from outside the field of view can reach the detector directly
as long as it avoids the secondary mirror. To prevent this most
telescopes have baffeling around the secondary mirror to reduce this
light path, but this reduction in the sky background comes at the cost
of vignetting mirror aperture. Many Cassegrain telescopes also have a
tube protruding from the center of the mirror which also cuts down on
the stray light, but with the same vignetting problems. Properly done,
the baffeling should eliminate any direct path to the sky. The Nasmyth
and prime foci also require baffling to prevent direct light from the
dome or mirror cell from reaching the detector. Even with this
baffling there exist paths where light can reach the detector after a
single bounce off of the various surfaces. 

In contrast the Schmidt telescope with a Newtonian focus has no direct
path for light to reach the detector, nor any single-bounce
paths. To further limit the possibility of stray light reaching the
detector the inside of the telescope tube has been covered with
light-absorbing flocking, a fuzzy velvet-like material.

Another upgrade made to the telescope in the past several years has
been the replacement of the Newtonian mirror and its mount. The
previous Newtonian mirror had been undersized and vignetted the field
of view. The mirror had also been supported by a structural element
attached to one side of the telescope tube. This element would flex
and cause the vignetting to change depending on the position of the
telescope. This changing illumination pattern posed a problem for
flat-fielding the detector and would have also affected the accuracy
of our reflection modeling. This support element was replaced with a
spider structure which is connected to four sides of the telescope
tube, thus reducing the flexure of the mirror. A properly-sized
Newtonian mirror was also installed at the same time.

\clearpage


%
%
%
%
\begin{table}
\begin{tabular}{l  c c c c}
\hline
\hline
Reflection Surface & Distance to surface & Reflectivity & Reflection Radius &
 Surface Brightness \\
& (cm) &  & (arcmin) & (mag $\mbox{arcsec}^{-2}$) \\
\hline
Dewar window bottom  & 0.64 & $0.4\%$ &  2.7   & 20.7 \\
Filter bottom & 3.3 & $1.6\%$ & 17.0   & 23.0 \\
Filter top    & 4.2 & $0.2\%$ & 19.5   & 25.2 \\
\hline
\end{tabular}
\caption{Summary of the reflections that are modeled and
  removed. These numbers use a CCD reflectivity averaged over the
  filter passband of 10\% (M. Lesser, private communication). The
  listed surface brightness is for a 450 second exposure of Arcturus
  ($M_V = -0.04$). \label{param-table}}
\end{table}

\begin{thebibliography}{}

\bibitem[Bernstein(2007)]{Bernstein07} Bernstein, R.~A.\ 2007, 
\apj, 666, 663 
\bibitem[de Jong(2008)]{deJong08} de Jong, R.~S.\ 2008, \mnras, 
388, 1521 
\bibitem[Feldmeier et al.(2002)]{Feldmeier02} Feldmeier, J.~J., 
Mihos, J.~C., Morrison, H.~L., Rodney, S.~A., \& Harding, P.\ 2002,
\apj, 575, 779 
\bibitem[Feldmeier et al.(2004)]{Feldmeier04} Feldmeier, J.~J., 
Mihos, J.~C., Morrison, H.~L., Harding, P., Kaib, N., 
\& Dubinski, J.\ 2004, \apj, 609, 617 

\bibitem[Gonzalez et al.(2005)]{Gonzalez05} Gonzalez, A.~H., 
Zabludoff, A.~I., \& Zaritsky, D.\ 2005, \apj, 618, 195 

\bibitem[Gordon et al.(1998)]{Gordon98} Gordon, K.~D., Witt, 
A.~N., \& Friedmann, B.~C.\ 1998, \apj, 498, 522 

\bibitem[Gregg \& West(1998)]{Gregg98} Gregg, M.~D., \&
West, M.~J.\ 1998, \nat, 396, 549


\bibitem[King(1971)]{King71} King, I.~R.\ 1971, \pasp, 83, 199 
\bibitem[McGaugh et al.(1995)]{McGaugh95} McGaugh, S.~S.,
Schombert, J.~M., \& Bothun, G.~D.\ 1995, \aj, 109, 2019
\bibitem[Malin \& Carter(1983)]{Malin83} Malin, D.~F., 
\& Carter, D.\ 1983, \apj, 274, 534 
\bibitem[Marshall(2004)]{Marshall04} Marshall, J.~J.\ 2004, 
Ph.D.~Thesis, 
\bibitem[Mart{\'{\i}}nez-Delgado et al.(2009)]{Delgado09} 
Mart{\'{\i}}nez-Delgado, D., Pohlen, M., Gabany, R.~J., Majewski, S.~R., 
Pe{\~n}arrubia, J., \& Palma, C.\ 2009, \apj, 692, 955 

\bibitem[Mihos et al.(2005)]{Mihos05} Mihos, J.~C., Harding, 
P., Feldmeier, J., \& Morrison, H.\ 2005, \apjl, 631, L41 
\bibitem[Morrison et al.(1997)]{Morrison97} Morrison, H.~L., 
Miller, E.~D., Harding, P., Stinebring, D.~R., 
\& Boroson, T.~A.\ 1997, \aj, 113, 2061 
\bibitem[Nassau(1945)]{Nassau45} Nassau, J.~J.\ 1945, \apj, 101, 
275 
\bibitem[Pohlen et al.(2002)]{Pohlen02} Pohlen, M., Dettmar, 
R.-J., L{\"u}tticke, R., \& Aronica, G.\ 2002, \aap, 392, 807 
\bibitem[Racine(1996)]{Racine96} Racine, R.\ 1996, \pasp, 108, 
699 
\bibitem[Sandage(1976)]{Sandage76} Sandage, A.\ 1976, \aj, 81, 
954 
\bibitem[Sprayberry et al.(1996)]{Sprayberry96} Sprayberry, D., 
Impey, C.~D., \& Irwin, M.~J.\ 1996, \apj, 463, 535 
\bibitem[Surma et al.(1990)]{Surma90} Surma, P., Seifert,
W., \& Bender, R.\ 1990, \aap, 238, 67

\bibitem[Uson et al.(1991)]{Uson91} Uson, J.~M., Boughn, 
S.~P., \& Kuhn, J.~R.\ 1991, \apj, 369, 46 

\bibitem[de Vaucouleurs et al.(1991)]{RC3} de Vaucouleurs, 
G., de Vaucouleurs, A., Corwin, H.~G., Jr., Buta, R.~J., Paturel, G., 
\& Fouque, P.\ 1991, Volume 1-3, XII, 2069 pp.~7 figs..~
Springer-Verlag Berlin Heidelberg New York,  

\bibitem[Witt et al.(2008)]{Witt08} Witt, A.~N., Mandel, S., 
Sell, P.~H., Dixon, T., \& Vijh, U.~P.\ 2008, \apj, 679, 497 



\end{thebibliography}
\end{document}